\documentclass[english,aps,rpm,reprint,onecolumn]{revtex4-2}
\usepackage[T1]{fontenc}
\usepackage[latin9]{inputenc}
\usepackage{geometry}
\usepackage{babel}
\usepackage{units}
\usepackage{amsmath}
\usepackage{amsthm}
\usepackage{amssymb}
\usepackage{stmaryrd}
\usepackage{esint}
\usepackage[unicode=true]
 {hyperref}

\makeatletter

\newcommand{\noun}[1]{\textsc{#1}}

\numberwithin{equation}{section}
\numberwithin{figure}{section}

\makeatother

\begin{document}
\title{A simple means for deriving quantum mechanics}
\author{Eric Tesse}
\email{eric.s.tesse@gmail.com}

\selectlanguage{english}%
\begin{abstract}
A type of mechanics will be presented that possesses some distinctive
properties. On the one hand, its physical description \& rules of
operation are readily comprehensible \& intuitively clear. On the
other, it fully satisfies all observable predictions of non-relativistic
quantum mechanics. Within it, particles exist at points in space,
follow continuous, piecewise differentiable paths, and their linear
momentum is equal to their mass times their velocity along their path.
Yet the probabilities for position and momentum, conditioned on the
state of the particle's environment, follow the rules of quantum theory.
Indeed, all observable consequences of quantum theory are satisfied;
particles can be entangled, have intrinsic spin, this spin is not
local to the particle, particle identity can effect probabilities,
and so forth. All the rules of quantum mechanics are obeyed, and all
arise in a straightforward fashion. After this is established, connections
will be drawn out between this type of mechanics and other types of
quantum worlds; those that obey Bohmian mechanics, stochastic mechanics,
the many worlds interpretation, and physical collapse. In the final
section, a relativistic version of the mechanics will be presented.
\end{abstract}
\maketitle

\section{Introduction}

It is often assumed that there does not exist any simple, intuitive
means for understanding quantum mechanics; any attempt to understand
it will be profoundly strange, and present deep conceptual challenges.
This view may not warranted.

In the pages that follow, a mechanical theory will be presented from
which quantum theory can be derived. Within any universe that obeys
this mechanical theory, all observations will be wholly consistent
with quantum theory. The mechanical theory is distinguished by both
the simplicity of its mathematical axioms, and the ease with which
one may conceptualize systems that satisfy the theory. It is sufficiently
simple to challenge the notion that quantum theory is inherently strange,
or presents any insurmountable conceptual difficulties.

Further, other types of worlds that satisfy quantum theory are closely
related to this simple one. These include worlds that obey Bohmian
mechanics, and those that obey the many worlds interpretation. It
is therefore possible that the simple mechanical theory may play a
central role in understanding the full range of ontologies that satisfy
quantum theory. 

\subsection*{Rudimentary concepts}

Before getting started, it will be helpful to introduce some basic
definitions. Consider a system that, at each moment in time, is in
some physical state. A ``system path'' is a function that maps time
to system states, one state at each time, such that the system is
capable of passing through all the path's states in sequence. If $\bar{p}$
is a path for a system, $\bar{p}(t)$ is the state at time $t$.

It will be assume that, if the system is in state $S$, everything
that is known at present about the past and the future is given by
the set of paths that pass through $S$. For example, the statement
``I know I took my keys this morning'' is true if and only if in
all world paths that pass through a state that contains my current
brain state, I took my keys in the morning of the current day. (Otherwise
``I'm pretty sure I took my keys this morning'' would be a more
accurate statement.)

Similarly, if $\bar{p}_{1}$ and $\bar{p}_{2}$ are paths of a closed
system, and $\bar{p}_{1}(t)=\bar{p}_{2}(t)$, then if the system is
in state $\bar{p}_{1}(t)$ at time $t$ there is no way to know if
the system reached this state via $\bar{p}_{1}$ or $\bar{p}_{2}.$
If there is a recording of which path was taken, then this recording
would be part of the state at time $t$, in which case $\bar{p}_{1}(t)\neq\bar{p}_{2}(t)$.

If $\bar{p}$ is a system path, define $\bar{p}[-\infty,t]$ to be
the semi-path up through time $t$ (that is, $\bar{p}[-\infty,t]$
is path $\bar{p}$ restricted to the domain $[-\infty,t]$), and $\bar{p}[t,\infty]$
to be the semi-path from time $t$ onward. If $\bar{p}_{1}$ and $\bar{p}_{2}$
are paths, and $\bar{p}_{1}(t)=\bar{p}_{2}(t)$, define $\bar{p}_{1}[-\infty,t]\circ\bar{p}_{2}[t,\infty]$
to be the function of time that's equal to $\bar{p}_{1}$ through
time $t$, and equal to $\bar{p}_{2}$ from time $t$ onward. This
operation is called ``concatenation''. $\bar{p}_{1}[-\infty,t]\circ\bar{p}_{2}[t,\infty]$
may not be an allowed system path, even if $\bar{p}_{1}$ and $\bar{p}_{2}$
are. If $\bar{p}_{1}[-\infty,t]\circ\bar{p}_{2}[t,\infty]$ is an
allowed system path for any pair of system paths, $\bar{p}_{1},\bar{p}_{2}$,
and any time, $t$, s.t. $\bar{p}_{1}(t)=\bar{p}_{2}(t)$, then the
system will be said to be ``closed under concatenation''.

\section{\label{sec:Introducing-positional-systems}Positional systems}

If you want to discuss something, you have to give it a name. We will
call the type of mechanics that's about to be described ``positional
mechanics'', and the systems that obey its rules ``positional systems''.

Positional systems are a large, easy to define class of physical systems
that display many of the characteristics of quantum mechanics. They
will be presented in this section. In the following section, the subclass
of positional systems that obey quantum theory in all detail will
be defined; elements of that subclass will be called ``quantum positional
systems''.

\subsection{Defining positional systems\label{subsec:Positional-Systems}}

Positional systems share many features with classical systems. They
are composed of particles and fields, and these particles \& fields
interact. Particles follow continuous, piecewise differentiable paths
through space, and a particle's linear momentum is equal to its mass
times its path's tangent vector. Unlike classical systems, it will
be seen that particle paths have constant velocity except at discrete
times, when the velocity can abruptly shift.

Particle states include position, but not velocity; the particle's
velocity is derived from its path (this is the source of the ``position''
in ``positional systems''). A particle's state also contains the
identifying characteristics of the particle, such as its mass and
charge. For a positional system as a whole, the particle portion of
its state is simply the set of the system's individual particle states,
one element in the set for each particle in the system. The full state
of a positional system will also include the state of the system's
interactive fields. 

As with classical fields, the positional field state carries momentum
and energy. In positional systems, particles \& fields interact non-deterministically
at discrete times; these interactions redistribute momentum and energy
among the particles and fields. The interactions will therefore cause
a particle's velocity to abruptly change. We can also expect the particle-field
interactions to result in the field state carrying information about
the full system path. Field states \& their paths will be more fully
developed when relativistic positional systems are described in Sec.
\ref{sec:Relativistic-Amendments}. For now, the few just mentioned
properties will be sufficient.

For a closed positional system, define the system's ``r-set'' (``reduced
set'') to be the set of all system paths that satisfy the following
condition: if, throughout any time interval, any subsystem does not
interact with its environment, then that subsystem's total momentum
is conserved throughout that time interval. Thus, in the r-set, the
momentum of a free particle is conserved, as is the momentum of the
closed system as a whole.

The final property of positional systems is that a closed system's
set of possible paths is the closure of its r-set under concatenation.
To be explicit: if $R_{0}$ is a closed system's r-set, define $R_{n+1}$
to be the smallest set s.t. for any $\bar{p}_{1},\bar{p}_{2}\in R_{n}$,
and any time $t$ s.t. $\bar{p}_{1}(t)=\bar{p}_{2}(t)$, $\bar{p}_{1}[-\infty,t]\circ\bar{p}_{2}[t,\infty]\in R_{n+1}$;
the set of possible paths is $\bigcup_{n\in\mathbb{N}}R_{n}$. Given
that all retained information about a closed system's past is encoded
in its state, it follows that direct observation cannot disclose whether
such a concatenation has taken place. 

For an individual particle, the only thing that can change when a
concatenation occurs is its velocity. As with field interactions,
upon a concatenation the particle velocity can experience an instantaneous
shift. If velocity were included in the particle state, then nothing
in the particle paths could change upon concatenation; this is why
velocity is excluded from the particle state.

As mentioned, the field state may have information about the particle's
velocity. If so, this will limit the the possible velocities that
the particle can have following a concatenation.

There is a useful, alternate way to view the closure of the r-set
under concatenation. As a closed system moves along an r-set path
it can, at any random time, select a new r-set path from among the
paths that share its current state. When a new path is chosen, the
change will be difficult to detect, because all aspects of the newly
chosen path will be consistent with all that is known about the prior
path.

In the following section it will be seen that these few simple assumptions
are sufficient to motivate many of the signature behaviors of quantum
systems. In Section \ref{sec:Probability-dynamics}, several parameters
will be tuned so as to ensure that quantum theory is fully satisfied.

\subsection{Characteristics of positional systems \label{sec:Characteristics-of-Positional}}

Before further developing positional systems, we briefly consider
how quantum behavior can arise from such a rudimentary construction.

\subsubsection{Non-determinism \& momentum conservation}

Positional systems are non-deterministic, yet no measurement can ever
reveal a violation of momentum conservation. Let's say that it is
known that a system's momentum was measured to be in range $R$, and
that that system has not interacted with its environment since that
measurement took place. For momentum conservation to be respected,
a momentum measurement performed now must be consistent with $R$. 

If the statement ``It is known that a system's momentum was measured
to be in range $R$'' is true, it must be the case that along any
path that passes through the current state of the system and its environment,
the system's momentum had been measured, and was found to be in range
$R$. The same goes for the statement ``It is known that the system's
has not interacted with its environment since the measurement''.
If the current state is $p$, and these things are known in state
$p$, they must be known in every state that the closed system can
pass through when evolving from the momentum measurement to state
$p$. This can be shown as follows: take any $\bar{p}[-\infty,t]$
s.t. $\bar{p}(t)=p$, and say that on this path the measurement took
place at time $t_{0}$. Take any $t_{0}<t^{\prime}<t$, and assign
$\bar{p}(t^{\prime})=p^{\prime}$. If there's a $\bar{p}^{\prime}[-\infty,t^{\prime}]$
s.t. $\bar{p}^{\prime}(t^{\prime})=p^{\prime}$ and no measurement
took place on $\bar{p}^{\prime}[-\infty,t^{\prime}]$, then by closure
under concatenation there would be a path passing through $p$ on
which the measurement does not take place. This violates our initial
assumption, so there can be no such $\bar{p}^{\prime}[-\infty,t^{\prime}]$.

So let's say that it is known that a system's momentum was measured
to be in range $R$, and that the system has not interacted with its
environment since that measurement took place. Along any r-set path,
the system's total momentum will be conserved, and so its momentum
will be in $R$. If the system has not followed an r-set path, then
there have been concatenations, which is equivalent to choosing a
succession of new r-set paths, each sharing the state of the existing
path at the time that the new path is chosen. For reasons stated above,
along any new r-set path, while the system may have a different momentum
than it had on the old path, the system's momentum was still measured
to be in $R$, and the system's momentum has been conserved since
the measurement took place. Upon concatenation, the system's momentum
must therefore continue to be in $R$. Even though positional particles
can non-deterministically change their velocities, no violation of
conservation of momentum can ever be detected.

\subsubsection{Non-locality}

Positional systems also exhibit non-locality of a type often associated
with quantum systems. This can be seen in the following simple example.
Consider a pair of free particles whose total momentum is known to
be zero, meaning that the pair's environment is in a state that can
only be reached when the pair's total momentum is zero. Upon concatenation,
the total momentum must then remain zero. This means that if, upon
concatenation, one particle's momentum changes from $\vec{p}_{1}$
to $\vec{p}_{2}$, the other particle's momentum must switch from
$-\vec{p}_{1}$ to $-\vec{p}_{2}$. These two free, non-interacting
particles are entangled.

\subsubsection{Non-additive probabilities}

Imagine a sequence of measurements where for the $i^{th}$ measurement
in the sequence, the possible outcomes are $\left\{ O_{1}^{i},O_{2}^{i},O_{3}^{i},...\right\} $.
Now imagine an identical sequence of measurements, except that in
this one, the possible outcomes of the $i^{th}$ measurement are $\left\{ O_{1}^{i}\bigcup O_{2}^{i},O_{3}^{i},...\right\} $.
In the first sequence, the experimental equipment will be in a different
state, depending on whether $O_{1}^{i}$ or $O_{2}^{i}$ occur. In
the second sequence, the environmental state will be the same for
$O_{1}^{i}$ or $O_{2}^{i}$. Therefore, in the second sequence, following
the $i^{th}$ measurement, a path that would have resulted in $O_{1}^{i}$
can be concatenated with a path that would have resulted $O_{2}^{i}$,
while in the first sequence, paths from the two outcomes can never
be concatenated. As a result, the set of possible paths are different
for these two measurement sequences. Under these conditions, not even
traditional probability theory would require additivity.

\subsubsection{Particle Identity}

Concatenation may also affect probabilities based on particle identity.
Consider a system consisting of two particles, and a pair of paths
for that system, $\bar{p}_{1}$ and $\bar{p}_{2}$, s.t. $\bar{p}_{1}(t)=\{(\vec{x}_{1},\iota_{1}),(\vec{x}_{2},\iota_{2})\}$
and $\bar{p}_{2}(t)=\{(\vec{x}_{1},\iota_{2}),(\vec{x}_{2},\iota_{1})\}$,
where $\iota_{1}$ and $\iota_{2}$ are the particle's identifying
characteristics, and $\vec{x}_{1}\neq\vec{x}_{2}$. These paths can
be concatenated at $t$ if and only if $\iota_{1}=\iota_{2}$. This
means that paths will often exist for collections of identical particles
that do not exist for collections of distinguishable particles. We
can therefore expect particle identity to effect probabilities.

\subsubsection{Uncertainty}

The field state will generally contain information about a particle's
position \& path. However, due to the discrete \& non-deterministic
nature of particle-field interactions, we do not expect this information
to be complete. In particular, for any particular particle, if we
are given the state of every other particle and the state of the fields,
we can expect to know something about that particle's position and/or
velocity, but we can not expect to know everything. It follows that
when performing a measurement on a physical system, given the complete
state of all equipment used to perform the measurement, we can never
expect complete knowledge of the measured system's state and/or instantaneous
path.

Relatedly, because concatenation is dependent on the state of the
closed system as a whole, the information stored in the field state
can effect the system's evolution.

\subsubsection{Further effects related to interaction: Tunneling, reflection, and
refraction}

Particles interact with fields in the manner described in Sec \ref{subsec:Positional-Systems}.
Because these interactions are discrete and non-deterministic, the
broad outlines of effects such as tunneling, reflection, and refraction
are relatively easy to motivate. 

Tunneling occurs when a particle passes through a thin barrier without
interacting with the barrier's fields. Similarly, the incident particle
may recoil from a barrier we would (classically) expect it to pass
through, if it receives an anomalously strong momentum transfer from
the fields. If the barrier has a hole in it, it is possible for a
particle to pass through the hole and still be deflected, because
the barrier's fields can non-deterministically kick a particle as
it passes through the hole. In general, the likelihood of these effects
will not be due to purely local factors; they will depend on the system
state as a whole. In this, and in their discrete nature, particle
interaction in positional systems are akin to concatenation. These
similarities between concatenation and particle interaction will be
utilized in the next section, where positional probability dynamics
will be taken up.

\subsection{Probability dynamics\label{sec:Probability-dynamics}}

\subsubsection{\label{subsec:Single-particle-time}Single particle time evolution}

We will now turn to the time evolution of positional system probabilities,
starting with the probabilities for a single particle within a positional
system. 

Define $\eta(\vec{x},\vec{v}|e)$ to be the probability that the particle
is at position $\vec{x}$ with velocity $\vec{v}$, given the the
particle's environment is in state $e$. $\eta(\vec{x},\vec{v}|e)$
encapsulates the environment's information about the particle's path
at time $t$. $\eta$ can change over time due to changes in the information
that the environment possesses about the particle, and due to particle
dynamics. Any changes to the environment's information will immediately
be reflected in $\eta$. However, even if the environment's information
does not change, $\eta$ generally still will, because the impact
of the information changes over time. To determine how $\eta$ changes
with time when the environment's information about the particle is
constant, we only need to know how a particle's velocity can change
with time.

Positional dynamics are piecewise deterministic: due to interactions
and/or concatenations, a particle's velocity will experience discrete,
non-deterministic jumps. Concatenations and interactions can be referred
to by the generic term ``momentum changing events''. Let's call
$Q_{c}(\vec{v}|\vec{x},\iota,e_{c})$ the probability that, upon concatenation,
the newly chosen particle path will have velocity $\vec{v}$, given
that state of the system as a whole is $(\vec{x},\iota,e_{c})$. Here,
$\vec{x}$ is the particle position, $\iota$ is the particle's identifying
characteristics, and $e_{c}$ are the aspects of the environmental
state that affect $Q_{c}$. Call the average number of concatenations
per unit time $\varGamma_{c}$. In general $\varGamma_{c}$ can be
a function of the state of the system as a whole.

Likewise, call the probability distribution of a particle's velocity
following an interaction $Q_{i}(\vec{v}|\vec{v}_{p},\vec{x},\iota,e_{i})$,
where $\vec{v}_{p}$ is the particle's velocity just prior to the
interaction, and $e_{i}$ are aspects of the environmental state that
can affect the interaction. If an interaction triggers a simultaneous
concatenation, then $Q_{i}$ incorporates both effects. Call the average
number of interactions per unit time $\varGamma_{i}$.

It will be assumed that a momentum changing event does not, in and
of itself, alter the likelihood that a later momentum changing event
will take place. It can only do so indirectly, by changing the state,
when $\varGamma_{c/i}$ is a function of the state. While the $\varGamma$'s
can depend on the state, they can be assumed to be smoothly varying.
For a small enough $\triangle t$, the probability for the number
of events of type $n$ ($n\in\{c,i\}$) in time $\triangle t$ is
then given by the Poisson distribution: $P_{n}(N)=\frac{1}{N!}(\varGamma_{n}\triangle t)^{N}e^{-\varGamma_{n}\triangle t}$.

If the information that the environment possess about the particle
path does not change due to interactions, we have for sufficiently
small $\triangle t$:

\begin{eqnarray*}
\eta(\vec{x},\vec{v},t) & = & e^{-(\varGamma_{c}+\varGamma_{i})\triangle t}\eta(\vec{x}-\vec{v}\triangle t,\vec{v},t_{0})+\\
 &  & \sum_{n\in\left\{ c,i\right\} }\varGamma_{n}e^{-\varGamma_{n}\triangle t}\intop_{0}^{\triangle t}ds\intop d\vec{v}_{0}\eta(\vec{x}_{0},\vec{v}_{0},t_{0})Q_{n}(\vec{v}|\vec{v}_{0},\vec{x}_{1},\iota,e_{n,t_{0}+s})+\\
 &  & \sum_{n,m\in\left\{ c,i\right\} }\varGamma_{n}\varGamma_{m}e^{-(\varGamma_{n}+\varGamma_{m})\triangle t}\intop_{0}^{\triangle t}ds_{2}\intop_{0}^{s_{2}}ds_{1}\intop d\vec{v}_{0}\intop d\vec{v}_{1}\eta(\vec{x}_{0}^{\prime},\vec{v}_{0},t_{0})\times\\
 &  & \;\;\;\;\;\;\;\;\;\;\;\;\;\;\;\;\;\;\;\;\;\;\;Q_{n}(\vec{v}_{1}|\vec{v}_{0},\vec{x}_{1}^{\prime},\iota,e_{n,t_{0}+s_{1}})Q_{m}(\vec{v}|\vec{v}_{1},\vec{x}_{2},\iota,e_{n,t_{0}+s_{2}})+...
\end{eqnarray*}

Where the first term corresponds to no momentum changing events between
times $t_{0}$ \& $t$, the second to one momentum changing event
during this time, etc. $\triangle t=t-t_{0}$; the second term variables
are $\vec{x}_{1}=\vec{x}-\vec{v}(\triangle t-s)$ \& $\vec{x}_{0}=\vec{x}_{1}-\vec{v}_{0}s$;
and the third term variables are $\vec{x}_{2}=\vec{x}-\vec{v}(\triangle t-s_{2})$,
$\vec{x}_{1}^{\prime}=\vec{x}_{2}-\vec{v}_{1}(s_{2}-s_{1})$, and
$\vec{x}_{0}^{\prime}=\vec{x}_{1}^{\prime}-\vec{v}_{0}s_{1}$. 

Take $\triangle t$ to be small enough so that only terms that are
first order in $\triangle t$ need to be considered. The above equation
becomes:

\begin{eqnarray*}
\eta(\vec{x},\vec{v},t_{0})+\triangle t\frac{\partial}{\partial t}\eta(\vec{x},\vec{v},t)|_{t_{0}} & = & \eta(\vec{x},\vec{v},t_{0})-\triangle t(\varGamma_{c}+\varGamma_{i})\eta(\vec{x},\vec{v},t_{0})-\triangle t\vec{v}_{0}\cdot\nabla\eta(\vec{x},\vec{v},t_{0}))+\\
 &  & \triangle t\intop d\vec{v}_{1}\eta(\vec{x},\vec{v}_{1},t_{0})\left(\varGamma_{c}Q_{c}(\vec{v}|\vec{x},\iota,e_{c,t_{0}})+\varGamma_{i}Q_{i}(\vec{v}|\vec{v}_{1},\vec{x},\iota,e_{i,t_{0}})\right)
\end{eqnarray*}

With $Q=\frac{\varGamma_{c}}{\varGamma_{c}+\varGamma_{i}}Q_{c}+\frac{\varGamma_{i}}{\varGamma_{c}+\varGamma_{i}}Q_{i}$
\& $\varGamma=\varGamma_{c}+\varGamma_{i}$, this yields:

\[
\frac{\partial}{\partial t}\eta(\vec{x},\vec{v},t)+\vec{v}\cdot\nabla\eta(\vec{x},\vec{v},t)=\varGamma\left(\rho(\vec{x},t)\left\langle Q(\vec{v}|\vec{v}_{b},\vec{x},\iota,e_{t})\right\rangle _{\vec{v}_{p}}-\eta(\vec{x},\vec{v},t)\right)
\]

Where $\rho(\vec{x},t)=\intop d\vec{v}\eta(\vec{x},t)$ is the particle's
position probability distribution, and $\left\langle Q(\vec{v}|\vec{v}_{p},\vec{x},\iota,e_{t})\right\rangle _{v_{p}}=\frac{1}{\rho(\vec{x},t)}\intop d\vec{v}_{p}Q(\vec{v}|\vec{v}_{p},\vec{x},\iota,e_{t})\eta(\vec{x},\vec{v}_{p},t)$
is the expectation value of $Q(\vec{v}|\vec{x},\iota,\vec{v}_{p},e_{t})$
when averaging over the prior velocity, $\vec{v}_{p}$. $e_{t}$ contains
those aspects of the environmental state that can affect either $Q_{c}$
or $Q_{i}$.

For enhanced readability, define $q(\vec{v}|\vec{x},\iota,e_{t})\equiv\left\langle Q(\vec{v}|\vec{v}_{b},\vec{x},\iota,e_{t})\right\rangle _{\vec{v}_{b}}$.
$q(\vec{v}|\vec{x},\iota,e_{t})$ is the probability distribution
for the outgoing velocity upon a momentum changing event, given the
state of the closed system. With it, we have the desired equation
of motion:

\begin{equation}
\frac{\partial}{\partial t}\eta(\vec{x},\vec{v},t)+\vec{v}\cdot\nabla\eta(\vec{x},\vec{v},t)=\varGamma\left[\rho(\vec{x},t)q(\vec{v}|\vec{x},\iota,e_{t})-\eta(\vec{x},\vec{v},t)\right]\label{eq:EOM}
\end{equation}

As previously mentioned, ``$e_{t}$'' need not be the full environmental
state. It need only refer to aspects of the environment that effect
$q$. For example, for concatenations, $q$ may only depend on the
information that the environment holds about the current particle
state. This information is captured by the position probability distribution,
$\rho(\vec{x},t)$, so in this case we'd have $q(\vec{v}|\vec{x},\iota,\rho(\vec{x},t))$.

\subsubsection{Solutions}

Let's assume that the environment's information about the particle
is constant, and that the last momentum changing event occurred at
time $t_{1}$. The current distribution is then 
\[
\eta(\vec{x},\vec{v},t)=\rho(\vec{x}-\vec{v}(t-t_{1}),t_{1})q_{c/i}(\vec{v}|\vec{x}-\vec{v}(t-t_{c}),\iota,e_{t_{1}})
\]
 where $q_{c}\equiv Q_{c}$ \& $q_{i}\equiv\left\langle Q_{i}\right\rangle _{v_{p}}$;
the $q$ to use depends on whether the event was a concatenation or
an interaction. The probability that the last momentum changing event
was a concatenation is $\frac{\varGamma_{c}}{\varGamma_{c}+\varGamma_{i}}$,
and the probability that it was an interaction is $\frac{\varGamma_{i}}{\varGamma_{c}+\varGamma_{i}}$.
The probability distribution for the time at which the last momentum
changing event occurred is $\tau(t_{1})=\varGamma e^{-\varGamma\left(t-t_{1}\right)}$.
Together these yield:

\begin{equation}
\eta(\vec{x},\vec{v},t)=\varGamma e^{-\varGamma t}\int_{0}^{t}e^{\varGamma t_{l}}\rho(\vec{x}-\vec{v}(t-t_{l}),t_{l})q(\vec{v}|\vec{x}-\vec{v}(t-t_{l}),\iota,e_{t_{c}})dt_{l}+e^{-\varGamma t}\eta(\vec{x}-\vec{v}t,\vec{v},0)\label{eq:Soln}
\end{equation}

where the final term is for no momentum changing events at all. Acting
$(\frac{\partial}{\partial t}-\vec{v}\cdot\nabla)$ on this expression
for $\eta(\vec{x},\vec{v},t)$ gives $\varGamma\rho(\vec{x},t)q(\vec{v}|\vec{x},\iota,e_{t})-\varGamma\eta(\vec{x},\vec{v},t)$,
showing that it is a solution to Eq.\ \ref{eq:EOM}.

If $\varGamma$ is large, we can expand Eq.\ \ref{eq:Soln} in powers
of $\frac{1}{\varGamma}$: 

\begin{align}
\eta(\vec{x},\vec{v},t)= & \rho(\vec{x},t)q(\vec{v}|\vec{x},\iota,e_{t})-\frac{1}{\varGamma}\left(\frac{\partial}{\partial t}+\vec{v}\cdot\nabla\right)\rho(\vec{x},t)q(\vec{v}|\vec{x},\iota,e_{t})+\label{eq: Soln L}\\
 & \;\frac{1}{\varGamma^{2}}\left(\frac{\partial}{\partial t}+\vec{v}\cdot\nabla\right)^{2}\rho(\vec{x},t)q(\vec{v}|\vec{x},\iota,e_{t})+...\nonumber 
\end{align}

Plugging this into $\varGamma[\rho(\vec{x},t)q(\vec{v}|\vec{x},\iota,e_{t})-\eta(\vec{x},\vec{v},t)]$
yields $(\frac{\partial}{\partial t}+\vec{v}\cdot\nabla)\eta(\vec{x},\vec{v},t)$,
showing that this too is a solution.

\subsubsection{Many particles}

It is relatively straightforward to repeat the above calculations
for multiparticle systems. They yield the multiparticle equation of
motion

\begin{align}
 & \frac{\partial}{\partial t}\eta(\vec{x}_{1},\vec{v}_{1},\iota_{1},\vec{x}_{2},\vec{v}_{2},\iota_{2},...;t)+\sum_{i}\vec{v}_{i}\cdot\nabla_{i}\eta(\vec{x}_{1},\vec{v}_{1},\iota_{1},\vec{x}_{2},\vec{v}_{2},\iota_{2},...;t)=\label{eq: MP EOM}\\
 & \;\varGamma[\rho(\vec{x}_{1},\iota_{1}\vec{x}_{2},\iota_{2},...;t)q(\vec{v}_{1},\vec{v}_{2},..|(\vec{x}_{1},\iota_{1}),(\vec{x}_{2},\iota),...,e_{t})-\eta(\vec{x}_{1},\vec{v}_{1},\iota_{1},\vec{x}_{2},\vec{v}_{2},\iota_{2},...;t)]\nonumber 
\end{align}

with the obvious multiparticle versions of Eqs. \ref{eq:Soln} \&
\ref{eq: Soln L}. By assumption, $\eta(\vec{x}_{1},\vec{v}_{1},\iota_{1},\vec{x}_{2},\vec{v}_{2},\iota_{2},...;t)$
is totally symmetric with regard to the interchange of any of the
$(\vec{x},\vec{v},\iota)$. It will often be the case that the number
of particles and their identifying characteristics are given and constant.
In such cases we can write $\eta(\vec{x}_{1},\vec{v}_{1},\iota_{1},\vec{x}_{2},\vec{v}_{2},\iota_{2},...;t)$
as $\eta_{\vec{\iota}}(\vec{x}_{1},\vec{v}_{1},\vec{x}_{2},\vec{v}_{2},...;t)$,
where $\vec{\iota}$ is the tuple of identifying characteristics:
the $i^{th}$ particle has characteristics $\iota_{i}$, and position
and velocity $(\vec{x}_{i},\vec{v}_{i})$.

\subsubsection{\label{subsec:Highly-frequent-concatenations}Highly frequent concatenations}

Let's consider the case in which $\varGamma_{i}$ \& $\varGamma_{c}$
are sufficiently large, and $q_{i}$ \& $q_{c}$ are sufficiently
smoothly varying, so that even for a fairly large number of momentum
changing events, $N$, $q_{i/c}$ are nearly constant over time period
$\frac{N}{\varGamma}$. For concreteness, one may assume that $\frac{1}{\varGamma}$
is on the order of the Planck time, $\sim10^{-43}sec$. 

Take $\rho(\vec{x}^{\prime},t^{\prime}|\vec{x},e,t)$ to be the particle
position probability distribution at time $t^{\prime}$, given that
at time $t<t^{\prime}$ the closed system as a whole was in state
$\left(\vec{x},e\right)$, $\vec{x}$ being the particle's position
at $t$. If $\varGamma$ and $q$ are nearly constant over the time
interval $[t,t^{\prime}]$ and $t^{\prime}-t\gg\frac{1}{\varGamma}$,
then due to the law of large numbers $\rho(\vec{x}^{\prime},t^{\prime}|\vec{x},e_{t},t)$'s
standard deviation will be small.

A brief proof of this assertion: For each $i<N$, define $\triangle\vec{x}_{i}$
be the difference in particle position between the $i^{th}$ and $(i+1)^{th}$
momentum changing event. Ignoring any particle movement between $t$
and $t^{\prime}$ that occurs before the first event or after the
last (which by assumption are small) we have $\vec{x}^{\prime}-\vec{x}=\triangle\vec{x}_{1}+\triangle\vec{x}_{2}+...+\triangle\vec{x}_{N}$.
This gives $Var(\vec{x}^{\prime}-\vec{x})=Var\left(\sum_{j\leq N}\triangle\vec{x}_{j}\right)=Var\left(\sum_{j\leq N}\vec{v}_{j}\triangle t_{j}\right)$.
The distribution of $\vec{v}$ is given by $q$, and the distribution
of $\triangle t$ is $P(\triangle t)=\varGamma e^{-\varGamma\triangle t}$.
Because $q$ and $\varGamma$ are constant, the $\vec{v}_{j}\triangle t_{j}$
are independent and identically distributed, and so $Var\left(\sum_{j\leq N}\vec{v}_{j}\triangle t_{j}\right)=N*Var\left(\vec{v}\triangle t\right)$.
Because $\vec{v}$ and $\triangle t$ are independent, $Var(\vec{v}\triangle t)=\left\langle \vec{v}^{2}\right\rangle \left\langle \triangle t^{2}\right\rangle -\left\langle \vec{v}\right\rangle ^{2}\left\langle \triangle t\right\rangle ^{2}=\frac{2}{\varGamma^{2}}\left\langle \vec{v}^{2}\right\rangle -\frac{1}{\varGamma^{2}}\left\langle \vec{v}\right\rangle ^{2}$.
This yields $Var(\vec{x}^{\prime}-\vec{x})=N*\left(\frac{2}{\varGamma^{2}}\left\langle \vec{v}^{2}\right\rangle -\frac{1}{\varGamma^{2}}\left\langle \vec{v}\right\rangle ^{2}\right)$.
Since $N\cong\varGamma\left(t^{\prime}-t\right)$, $Var(\vec{x}^{\prime}-\vec{x})=\frac{(t^{\prime}-t)}{\varGamma}\left(2\left\langle \vec{v}^{2}\right\rangle -\left\langle \vec{v}\right\rangle ^{2}\right)$.
If $\varGamma$ is large, this is vanishingly small.

Define $\vec{\nu}(\vec{x},t)\equiv\frac{1}{\rho(\vec{x},t)}\int\vec{v}\eta(\vec{x},\vec{v},t)d^{3}v$;
$\vec{\nu}(\vec{x},t)$ to be the velocity expectation as a function
of position. For time periods much greater than $\frac{1}{\varGamma}$,
but small enough for $q$ to be roughly constant, by the above proof
it is highly likely that $\frac{\triangle\vec{x}}{\triangle t}=\vec{\nu}(\vec{x},t)$.
In other words, for large $\varGamma$, particle paths almost always
zigzag rapidly around their mean velocity. In this regime, the distribution
of particle paths at time $t$ can be well described by $\rho(\vec{x},t)$
and $\vec{\nu}(\vec{x},t)$ rather than the more complete description
given by $\eta(\vec{x},\vec{v},t)$. The equations of motion for $\rho(\vec{x},t)$
and $\vec{\nu}(\vec{x},t)$ follow from equation \ref{eq:EOM}: 
\begin{equation}
\frac{\partial}{\partial t}\rho+\nabla\cdot(\rho\vec{\nu})=0\label{eq:cont}
\end{equation}
 and 
\begin{equation}
\rho(\frac{\partial}{\partial t}\vec{\nu}+\vec{\nu}\cdot\nabla\vec{\nu})+\nabla\cdot\overset{\leftrightarrow}{\Lambda}=\vec{\alpha}\label{eq:navier}
\end{equation}
where $\overset{\leftrightarrow}{\Lambda}=\int\left(\vec{v}-\vec{\nu}\right)\left(\vec{v}-\vec{\nu}\right)\eta d^{3}v$
is the local velocity variance (together with a factor of $\rho$),
and $\vec{\alpha}=\varGamma\rho\left(\intop\vec{v}qd\vec{v}-\vec{\nu}\right)$.

No matter how large $\varGamma$ may be, all the properties of Sec
\ref{subsec:Positional-Systems} and behavioral characteristics of
Sec \ref{sec:Characteristics-of-Positional} continue to hold. Similarly,
no matter how large $\varGamma$ may be, the particle's instantaneous
velocity is unlikely to be $\vec{\nu}(\vec{x},t)$. However, for time
scales that are large compared to $\frac{1}{\varGamma}$, but over
which the change in $\vec{\nu}(\vec{x},t)$ is negligible, the average
velocity is very likely to be $\vec{\nu}(\vec{x},t)$. 

For multiparticle systems, define $\vec{\nu}_{\vec{\iota},i,}=\frac{1}{\rho_{\vec{\iota}}}\int\vec{v}_{i}\eta_{\vec{\iota}}d^{3}v_{1}...d^{3}v_{N}$,
$\overset{\leftrightarrow}{\Lambda}_{\vec{\iota},ij}=\int\left(\vec{v}_{i}-\vec{\nu}_{i}\right)\left(\vec{v}_{j}-\vec{\nu}_{j}\right)\eta_{\vec{\iota}}d^{3}v_{1}...d^{3}v_{N}$,
and $\vec{\alpha}_{\vec{\iota},i}=\varGamma\rho_{\vec{\iota}}\left(\int\vec{v}_{i}qd^{3}v_{1}...d^{3}v_{N}-\vec{\nu}_{\vec{\iota},i}\right)$.
Eqs.\ \ref{eq:cont} \& \ref{eq:navier} then become

\begin{equation}
\frac{\partial}{\partial t}\rho_{\vec{\iota}}+\nabla\cdot(\rho_{\vec{\iota}}\sum_{k}\vec{\nu}_{\vec{\iota},k})=0\label{eq: mp cont}
\end{equation}

\begin{equation}
\rho_{\vec{\iota}}\left(\frac{\partial}{\partial t}\vec{\nu}_{\vec{\iota}i}+\sum_{k}\vec{\nu}_{\vec{\iota}k}\cdot\nabla_{k}\vec{\nu}_{\vec{\iota}i}\right)+\sum_{k}\nabla_{k}\cdot\overset{\leftrightarrow}{\Lambda}_{\vec{\iota},ki}=\vec{\alpha}_{\vec{\iota},i}\label{eq: mp navier}
\end{equation}

where the ``$i$'' subscript refers to the $i^{th}$ particle. As
these are of the form of the Madelung equations, we appear to be approaching
quantum theory.

\section{Quantum positional systems \label{sec:Quantum-positional-systems}}

\subsection{Schrodinger distributions\label{subsec:Schrodinger-systems}}

Let's take $\varGamma$ to be large, and tune the parameters in Eq.\ \ref{eq:navier}
so that $\rho$ and $\vec{\nu}$ conform to quantum theory. $(\rho,\vec{\nu})$
that behave like solutions to the Schrodinger equation will be termed
``Schrodinger distributions''. Schrodinger distributions are not
the only type of distribution for quantum positional systems; others
will be discussed in Sec \ref{subsec:Other dists}. They are, however,
a uniquely important class of distributions, as the other types can
be defined in terms of them.

In this section, spin will be ignored. Spin will be incorporated in
Sec \ref{subsec:Spin}.

Start with Eq.\ \ref{eq:navier}, $\rho(\frac{\partial}{\partial t}\vec{\nu}+\vec{\nu}\cdot\nabla\vec{\nu})+\nabla\cdot\overset{\leftrightarrow}{\Lambda}=\vec{\alpha}$.
We will take $\vec{\alpha}$ to be of the form $\vec{\alpha}=\frac{\rho}{m}\vec{F}+(\frac{\hbar}{2m})^{2}\nabla(\nabla^{2}\rho)+...$.
The first term, $\frac{\rho}{m}\vec{F}$, is due to interaction with
the environment while the second, $(\frac{\hbar}{2m})^{2}\nabla(\nabla^{2}\rho)$,
may be assumed to be due to concatenation. ($\rho$ is an artifact
of the environmental state, it is information the environment has
about the particle's position, and those aspects of the environmental
state may effect concatenation.)

$\overset{\leftrightarrow}{\Lambda}=\int(\vec{v}-\vec{\nu})(\vec{v}-\vec{\nu})\eta d^{3}v$
is the local velocity variance, meaning that $m^{2}\int\overset{\leftrightarrow}{\Lambda}d^{3}x$
is the total momentum variance. For a Schrodinger wave function, the
momentum variance is $(\frac{\hbar}{2})^{2}\int\rho(\frac{1}{\rho}\nabla\rho)(\frac{1}{\rho}\nabla\rho)d^{3}x$,
which requires $m^{2}\int\overset{\leftrightarrow}{\Lambda}d^{3}x=(\frac{\hbar}{2})^{2}\int\rho(\frac{1}{\rho}\nabla\rho)(\frac{1}{\rho}\nabla\rho)d^{3}x$.
Therefore $\overset{\leftrightarrow}{\Lambda}=(\frac{\hbar}{2m})^{2}\frac{1}{\rho}(\nabla\rho)(\nabla\rho)+\overset{\leftrightarrow}{Z}$,
where $\overset{\leftrightarrow}{Z}$ is some symmetric, second rank
tensor s.t. $\int\overset{\leftrightarrow}{Z}d^{3}x=0$ and $Z_{ii}\geq-(\frac{\hbar}{2m})^{2}\frac{1}{\rho}(\frac{\partial}{\partial x_{i}}\rho)^{2}$.
For Schrodinger distributions, $\vec{\alpha}=$ $\frac{\rho}{m}\vec{F}+(\frac{\hbar}{2m})^{2}\nabla(\nabla^{2}\rho)+\nabla\cdot\overset{\leftrightarrow}{Z}$.
Plugging this form of $\vec{\alpha}$ \& $\overset{\leftrightarrow}{\Lambda}$
into Eq. \ref{eq:navier} gives:

\begin{equation}
m\rho\left(\frac{\partial}{\partial t}\vec{\nu}+\vec{\nu}\cdot\nabla\vec{\nu}\right)=\rho\vec{F}+\frac{\hbar^{2}}{4m}\nabla\cdot\left(\nabla\nabla\rho-\frac{1}{\rho}(\nabla\rho)(\nabla\rho)\right)\label{eq: quant navier}
\end{equation}

Generalizing to multiple particle systems yields:

\begin{equation}
m\rho\left(\frac{\partial}{\partial t}\vec{\nu}_{\vec{\iota},i}+\sum_{k}\vec{\nu}_{\vec{\iota},k}\cdot\nabla_{k}\vec{\nu}_{\vec{\iota},i}\right)=\rho\vec{F_{i}}+\sum_{k}\frac{\hbar^{2}}{4m_{k}}\nabla_{k}\cdot\left(\nabla_{k}\nabla_{i}\rho-\frac{1}{\rho}(\nabla_{k}\rho)(\nabla_{i}\rho)\right)\label{eq: mp quant navier}
\end{equation}

where the ``$i$'' subscript refers to the $i^{th}$ particle.

Equations \ref{eq: quant navier}/\ref{eq: mp quant navier} \& \ref{eq:cont}/\ref{eq: mp cont}
are the Madelung equations\citep{Madelung}. They can be used to derive
the Schrodinger equation. To do so, start by noting that $\nabla\cdot\left(\nabla\nabla\rho-\frac{1}{\rho}(\nabla\rho)(\nabla\rho)\right)=-2\rho\nabla\left(\frac{1}{\sqrt{\rho}}\nabla^{2}\sqrt{\rho}\right)$,
and $\vec{\nu}\cdot\nabla\vec{\nu}=\frac{1}{2}\nabla\left(\vec{\nu}\cdot\vec{\nu}\right)$.
If $\vec{F}$ is conservative, $\vec{F}=-\nabla V$, we then have:
$m\frac{\partial}{\partial t}\vec{\nu}=-\nabla\left(\frac{m}{2}\vec{\nu}\cdot\vec{\nu}+V+\frac{\hbar^{2}}{2m}\frac{1}{\sqrt{\rho}}\nabla^{2}\sqrt{\rho}\right)$.
For conservative forces, $\frac{\partial}{\partial t}\vec{\nu}$ is
therefore a gradient field, and so for some $\vec{K}$, $m\vec{\nu}=\nabla S$+$\vec{K}$,
where $\vec{K}$ is a constant with respect to time. For Schrodinger
distributions, $\vec{K}$ is also a gradient. Assigning $m\vec{\nu}=\nabla\phi$,
this yields: 

\begin{equation}
\frac{\partial}{\partial t}\phi+\frac{1}{2m}\left(\nabla\phi\cdot\nabla\phi\right)+V+\frac{\hbar^{2}}{2m}\frac{1}{\sqrt{\rho}}\nabla^{2}\sqrt{\rho}=0\label{eq: quant h-j}
\end{equation}

For multiparticle systems, this is:
\begin{equation}
\frac{\partial}{\partial t}\phi_{\vec{\iota}}+\sum_{i}\frac{\hbar}{2m_{i}}(\nabla_{i}\phi_{\vec{\iota}})^{2}+V+\sum_{i}\frac{\hbar}{2m_{i}}\frac{1}{\sqrt{\rho_{\vec{\iota}}}}\nabla_{i}^{2}\sqrt{\rho_{\vec{\iota}}}=0\label{eq: mp quant h-j}
\end{equation}

From here on out, we will suppress the $\vec{\iota}$ subscript. For
either single or multiparticle distributions, define $\psi=\sqrt{\rho}e^{\frac{i}{\hbar}\phi}$.
If $\psi$ satisfies the Schrodinger equation then $(\rho,\phi)$
satisfies Eqs.\ \ref{eq:cont} \& \ref{eq: quant h-j} (or their
multiparticle counterparts, Eqs.\ \ref{eq: mp cont} \& \ref{eq: mp quant h-j})
. This can be seen by noting that Eq.\ \ref{eq: quant h-j} (\ref{eq: mp quant h-j})
is the real part of the Schrodinger equation, while Eq.\ \ref{eq:cont}
(\ref{eq: mp cont}) is the imaginary part.

To go the other way and ensure that $(\rho,\vec{j}=\frac{\rho}{m}\nabla\phi)$
satisfies Eqs.\ \ref{eq:cont} \& \ref{eq: quant h-j} if and only
if $\psi$ satisfies the Schrodinger equation, a final restriction
will have to be enforced: $e^{\frac{i}{\hbar}\phi}$ must be continuous.
With that, a positional system's $(\rho,\vec{j})$ is a Schrodinger
distribution of if and only if $\sqrt{\rho}e^{\frac{i}{\hbar}\phi}$
is a Schrodinger wavefunction.

If the force is not conservative, Eq.\ \ref{eq: quant navier} continues
to hold, but Eq.\ \ref{eq: quant h-j} generally will not. To generate
a version of Eq.\ \ref{eq: quant h-j} that is equivalent to Eq.\ \ref{eq: quant navier}
in the presence of a magnetic field, substitute $\nabla\phi\rightarrow\nabla\phi-q\vec{A}$,
where $q$ is the particle charge, and $\vec{A}$ is the vector potential.
For a multiparticle system, substitute $\nabla_{i}\phi\rightarrow\nabla_{i}\phi-q_{i}\vec{A}.$

For identical particles, $\rho$ is restricted to be zero whenever
two identical fermions are at the same point; this is the only restriction
on $\rho$. By the nature of $\eta$, $\rho$ will always be symmetric
for identical particles. The lion's share of Bose \& Fermi statistics
are due to $\vec{\nu}$. 

For Schrodinger distributions, $\rho$ can be any probability distribution
(with the exception of the mild restriction on identical fermions).
The restrictions are on $\vec{\nu}$. When $\varGamma$ is large,
$\vec{\nu}(\vec{x},t)=\int\vec{v}q(\vec{v}|\vec{x},\iota,e_{t})d^{3}v$,
so these restrictions are due to the nature of $q(\vec{v}|\vec{x},\iota,e_{t})$.

Given how $\psi$ has been derived from $\eta$, one would not expect
all information in $\eta$ to also be present in $\psi$. That is,
while $\psi$ can be derived from $\eta$, we do not expect to be
able to derive $\eta$ from $\psi$. Given what is known of the Wigner
quasiprobability distribution\citep{Wigner} \& related distributions,
we can conclude that $\eta$ can not be derived from $\psi$. In particular,
in the local momentum variance, $\int(\vec{v}-\vec{\nu})(\vec{v}-\vec{\nu})\eta d^{3}v=(\frac{\hbar}{2m})^{2}\frac{1}{\rho}(\nabla\rho)(\nabla\rho)+\overset{\leftrightarrow}{Z}$,
$\overset{\leftrightarrow}{Z}$ can not be written in terms of $\psi$
alone (or, equivalently, in terms $(\rho,\vec{\nu})$). 

\subsubsection*{Cycling back}

When describing the characteristics of positional systems in Sec \ref{sec:Probability-dynamics},
some of the characteristics depended on the probabilities of the outgoing
r-set paths following a momentum changing event. That is, they depended
on the nature of $q(\vec{v}|\vec{x},\iota,e_{t})$. These characteristics
include non-additivity, interference, and the effects of particle
identity, as well as interactive characteristics like tunneling \&
refraction.

For positional systems with large $\Gamma$, $q$ appears in $\vec{\nu}=\int\vec{v}qd\vec{v}$,
$\overset{\leftrightarrow}{\Lambda}=\rho\int\vec{v}\vec{v}qd^{3}v-\rho\vec{\nu}\vec{\nu}$,
and $\vec{\alpha}$. As is likely obvious, for any $q(\vec{v}|\vec{x},\iota,e_{t})$
that satisfies the above mentioned quantum positional system requirements
on $\vec{\nu}$, $\overset{\leftrightarrow}{\Lambda}$, and $\vec{\alpha}$,
the characteristics discussed in Sec \ref{sec:Probability-dynamics}
will appear in precisely the manner seen quantum systems.

\subsection{Observables }

\subsubsection{Position}

Within both positional systems and quantum theory, ``position''
is assigned its everyday meaning. Among observables, position is particularly
significant because the measurement of any other observable can be
translated into position; for example, by registering the outcome
as the position of a pointer in a gauge. This means that all other
observables can be defined operationally in terms of how such position
outcomes are produced. As J. S. Bell put it: ``in physics, the only
observations we must consider are position observations, if only the
positions of instrument pointers. {[}...{]} If you make axioms, rather
than definitions and theorems, about the measurement of anything else,
then you commit redundancy and risk inconsistency'' \citep{Bell 1}. 

In the prior section, it was established that quantum positional systems
reproduce quantum predictions for position. This means that, when
considering the experimental setup as a whole, quantum positional
systems will agree with quantum theory on all experimental outcomes.
It therefore agrees with quantum mechanics on all measurement predictions,
no matter the quantity being measured.

What remains of interest is the nature of these observable quantities
within positional systems. This will now be explored.

\subsubsection{Momentum}

Positional systems make a commitment to the nature of linear momentum.
It needs to be establish that this commitment is consistent with momentum
as it appears within quantum theory.

Let's start with a simple property of free positional particles. Let's
say it is known that at some time, $t_{0}$, a free positional particle
was in some bounded region of space, $R$. At any later time, $t_{1}$,
the particle's velocity must equal the velocity of some free particle
r-set path, so for some $\vec{x}_{0}\in R$, $\vec{v}(t_{1})=\frac{\vec{x}(t_{1})-\vec{x}_{0}}{t_{i}-t_{0}}$.
As the particle's velocity changes over time due to concatenations,
the $\vec{x}_{0}\in R$ in this equation can change, but one must
always exist. Therefore, by positional dynamics, if $R$ is bounded
and the particle is free, then as $t$ grows large it becomes increasingly
necessary that the particle's velocity at $\left(\vec{x},t\right)$
is near $\frac{\vec{x}(t)}{t}$; in the limit as $t\rightarrow\infty$,
the identity holds exactly. This means $\lim_{t\rightarrow\infty}\vec{v}(t)=\lim_{t\rightarrow\infty}\frac{\vec{x}(t)}{t}$. 

It follows that in the limit as $t\rightarrow\infty$ the probability
distribution for velocity, $\upsilon$, satisfies $\upsilon(\vec{v},t)d^{3}v=\rho(\vec{v}t,t)d^{3}(vt)$.
That is, $\lim_{t\rightarrow\infty}\upsilon(\vec{v},t)=\lim_{t\rightarrow\infty}t^{3}\rho(\vec{v}t,t)$.
For a free particle the velocity distribution is constant, so $\upsilon(\vec{v})=\lim_{t\rightarrow\infty}t^{3}\rho(\vec{v}t,t)$.
Switching to the momentum, $\vec{p}=m\vec{v}$, the momentum distribution,
$\mu$, is $\mu(\vec{p})=\lim_{t\rightarrow\infty}\left(\frac{t}{m}\right)^{3}\rho(\frac{1}{m}\vec{p}t,t)$.
For a Schrodinger distribution, $\rho(\vec{x},t)=\left|\psi(\vec{x},t)\right|^{2}$,
so $\mu(\vec{p})=\lim_{t\rightarrow\infty}\left(\frac{t}{m}\right)^{3}\left|\psi(\frac{1}{m}\vec{p}t,t)\right|^{2}$.

In Shucker \citep{Shucker}, and in G. C. Ghirard, C. Omero, A. Rimini,
and I. Weber \citep{Ghirard}, it is shown that if $\psi(\vec{x},t)$
is a solution of the free Schrodinger equation, $\int\left|\psi(\vec{x},t_{0})\right|^{2}d^{3}x=1$,
and $\int\left|\psi(\vec{x},t_{0})\right|d^{3}x<\infty$, then $\lim_{t\rightarrow\infty}\left(\frac{t}{m}\right)^{3}\left|\psi(\frac{1}{m}\vec{p}t,t)\right|^{2}=\left|\tilde{\psi}(\vec{p})\right|^{2}$,
where $\tilde{\psi}(\vec{p})$ is the Fourier transform of $\psi(\vec{x},t_{0})$.
If $\int\left|\psi(\vec{x},t_{0})\right|d^{3}x$ diverges, then the
equality holds almost everywhere: $\lim_{t\rightarrow\infty}\int\left|\left(\frac{t}{m}\right)^{3}\left|\psi(\frac{1}{m}\vec{p}t,t)\right|^{2}-\left|\tilde{\psi}(\vec{p})\right|^{2}\right|d^{3}p=0$.

For a free positional particle with a Schrodinger distribution, the
particle's momentum distribution is therefore $\left|\tilde{\psi}(\vec{p})\right|^{2}$.
This has been proved when the particle is in any arbitrarily large
region $R$. Since $R$ is arbitrarily large, it may be assumed to
hold in general. If the particle is not free, but is subject to conservative
forces, then the forces can be turned off without affecting the particle's
instantaneous velocity distribution, after which the above analysis
can be applied. The momentum distribution of a quantum positional
particle subject to a conservative force is therefore $\left|\tilde{\psi}(\vec{p},t)\right|^{2}$,
in agreement with quantum theory.

This leaves magnetic fields, which can be handled using the same method:
turn off the magnetic fields \& apply the above equality. Magnetic
fields have a force term of the form $-q\frac{\partial}{\partial t}\vec{A}$,
meaning that if we were to instantaneously turn off the fields, the
velocity will instantaneously shift from $\vec{v}\rightarrow\vec{v}+\frac{q}{m}\vec{A}$.
The probability distribution $\left|\tilde{\psi}(\vec{p},t)\right|^{2}$
is therefore on $\vec{p}=m\vec{v}+q\vec{A}$. For quantum positional
systems, a particle's quantum momentum can be identified as its conjugate
momentum.

\subsubsection{Spin \label{subsec:Spin} }

To this point, two aspects of a particle's environment effect $q(\vec{v}|\vec{x},\iota,e_{t})$:
those that generate forces on the particle, and those that contain
the information about the particle that yields $\rho$. There is one
further aspect of a particle's environmental state that effects $q(\vec{v}|\vec{x},\iota,e_{t})$,
and that is those aspects that specify the particle's spin. That spin
is an artifact of the particle's environment, and not the particle
itself, is indicated by the fact that it is an element of the wave
function; it is also indicated by Bell's theorem and the Kochen-Specker
theorem \citep{Bell 2,Kochen}.

All matter with which we are associating positional paths is composed
of spin $\nicefrac{1}{2}$ particles. When spin $\nicefrac{1}{2}$
particles are described by a Schrodinger distribution (as opposed
to more general density matrix distributions, to be discussed shortly),
the spin state can always be represented by a vector function of position,
$\vec{s}(\vec{x})$. $\vec{s}(\vec{x})$ is constrained to be a unit
vector at every point in space where $\rho$ is non-zero. $\vec{s}(\vec{x})$
has many properties in common with $\rho$, but differs in that $\vec{s}(\vec{x})$
does not carry information about the particle's current path. In particular,
spin is not some bit of information derived from $\eta(\vec{x},\vec{v})$
that has been left over after extracting $\rho$ and $\vec{\nu}$.
To see this, note that a particle with distribution $\eta(\vec{x},\vec{v})=\delta(\vec{v}-\vec{v}^{\prime})$
(homogeneous in space) can have any spin orientation.

With the inclusion of spin, the equation of motion for $(\rho,\vec{\nu})$
has to be expanded to encompass $(\rho,\vec{\nu},\vec{s})$. These
changes will be briefly described.

Spin does not alter the equation of continuity, Eq.\ \ref{eq:cont},
but the equation of motion for $\vec{\nu}$, Eq.\ \ref{eq: quant navier},
is effected. The right hand side gains an additional term of $-\frac{\hbar^{2}}{4m}\nabla\cdot\rho(\vec{x})\left[\left(\nabla\theta(\vec{x})\right)\left(\nabla\theta(\vec{x})\right)+\sin^{2}\theta(\vec{x})\left(\nabla\phi(\vec{x})\right)\left(\nabla\phi(\vec{x})\right)\right]$,
where $(\theta,\phi)$ is the spin direction. These additional terms
come from the velocity variance term in Eq.\ \ref{eq:navier}. If
the spin direction does not change with position, these terms are
zero.

In the absence of an external magnetic field, and in the absence of
any change in the information the environment has about the particle,
the equation of motion for $\vec{\sigma}(\vec{x})$ is: 
\begin{equation}
\frac{\partial}{\partial t}\vec{s}=-\vec{\nu}\cdot\nabla\vec{s}-\frac{\hbar}{2m\rho}\nabla\cdot\left[\rho\left(\nabla\vec{s}\right)\times\vec{s}\right]\label{eq:spin eom-1}
\end{equation}
The first term on the right hand side is due to the spin being carried
on the probability current. The second term represents torque generated
by gradients in the spin field. For this term, the ``$\mathbf{\cdot}$''
is on the two ``$\nabla$'', while the ``$\times$'' is on the
two ``$\vec{s}$''. It can perhaps be written more clearly as: $-\frac{\hbar}{2m\rho}\partial^{i}\left(\rho\left(\partial_{i}\vec{s}\right)\times\vec{s}\right)$.

In the presence of a magnetic field, add $\frac{\hbar}{2}\mu\nabla\left(\vec{B}\cdot\vec{s}\right)$
to the force term in Eq.\ \ref{eq: quant navier}, and $\frac{\hbar}{2}\mu\vec{s}\times\vec{B}$
to the right hand side of Eq.\ \ref{eq:spin eom-1}, where $\mu$
is the magnetic moment, $\frac{gq}{2m}$. $\frac{\hbar}{2}\mu\vec{s}\times\vec{B}$
generates Larmor precession on the spin, while the $\frac{\hbar}{2}\mu\nabla\left(\vec{B}\cdot\vec{s}\right)$
force is crucial for the Stern-Gerlach experiment. 

Finally, the restrictions on $\vec{\nu}$ described in Sec. \ref{subsec:Schrodinger-systems}
are now on $\vec{\nu}+\frac{\hbar}{2m}\cos\theta\nabla\phi$, where
$\left(\theta,\phi\right)$ is the spin direction. $\left(\rho,\vec{\nu},\vec{s}\right)$
can be encoded into a wavefunction as $\psi=\xi\sqrt{\rho}e^{\frac{i}{\hbar}\varphi}$,
where $\xi$ is the eigenspinor for the spin direction at each point
$\vec{x}$, and $\rho\vec{\nu}=\frac{\hbar}{2mi}\left(\psi^{\dagger}\nabla\psi-(\nabla\psi^{\dagger})\psi\right)$.

The majority of these terms are dynamic effects that occur when spin
changes with position. Within any region of space where the spin is
constant, only the magnetic field terms appear.

As can be seen, the spin value is a function of the possible particle
states (literally; $\vec{s}$ is a function of $\vec{x}$, restricted
to values where $\rho$ is non-zero), its evolution is effected by
the environment's information about the particle's path encapsulated
in $(\rho,\vec{\nu})$, and it value effects the particle's path.
These properties enable the spin to be determined by measurements
on the particle.

\subsubsection{Orbital Angular Momentum}

Let's assume a positional particle of mass $\mu$ has a static Schrodinger
distribution with a z-component of (quantum) angular momentum of $m\hbar$.
For simplicity, we'll assume $\nu_{z}=0$. If the particle is at point
$(r,\theta,z)$, its average velocity is $\vec{\nu}=\frac{m\hbar}{\mu r}\hat{\theta}$.
If $m$ is $0$, the particle executes jittery, Brownian type motion
centered at $(r,\theta,z)$. The jitter is responsible for the linear
momentum spectrum. If $m\neq0$, the average velocity describes a
circle passing through $(r,\theta,z)$ around the z-axis. The classical
angular momentum of this average velocity path is $m\hbar$, the same
as the quantum angular momentum. It is highly likely that the particle's
actual path will dart rapidly around the circle formed by the average
velocity.

Based on the nature of angular momentum distributions, quantum orbital
angular momentum does not measure any instantaneous aspect of the
particle path. A measurement of orbital angular momentum unfolds over
time. More saliently, the measurement changes the environmental state
so as to cause the jump probability, $q(\vec{v}|\vec{x},\iota,e)$,
to, on average, move the particle around in a circle. In other words,
$q(\vec{v}|\vec{x},\iota,e)$ is rotationally invariant. These are
the only rotationally invariant $q(\vec{v}|\vec{x},\iota,e)$. This
shares something in common with a spin measurement, in that a crucial
element of the measurement is the effect of the measurement on the
particle's environmental state, which in turn effects $q(\vec{v}|\vec{x},\iota,e)$.

\subsection{Other types of distributions\label{subsec:Other dists}}

In addition to Schrodinger distributions, quantum positional systems
support two further types of distributions: conditional distributions
and density matrix distributions. For ease of reading, in what follows
the ``$\vec{\iota}$'' subscript for identifying particle characteristics
will be suppressed, and the $\vec{\nu}$ \& $\vec{j}\equiv\rho\vec{\nu}$
vectors will span all components of all particles; if the system contains
$N$ particles these vectors have $3N$ components. 

Let's say that when some system is conditioned on the state of its
environment, it can be described by Schrodinger distribution. If we
were to further condition on some aspect of the system state, the
result is a conditional distribution. That is, assume that $\eta(...|e_{t})$
yields Schrodinger distribution $(\rho,\vec{\nu})$. Take some aspect
of the system state, $C(t)$, to be given. As with the environmental
state, $e_{t}$, we do not demand that something external ''knows''
$C(t)$, $C(t)$ it is simply asserted. The resulting $(\rho_{C(t)},\vec{\nu}_{C(t)})$
is a conditional distribution.

As an example, assume that $C(t)$ only asserts something about the
particle states. The conditional distribution is then $\rho_{C(t)}=\frac{1}{\int\rho|_{C(t)}}\rho|_{C(t)}$
and $\vec{\nu}_{C(t)}=\vec{\nu}|_{C(t)}$, where $\rho|_{C(t)}$ \&
$\vec{\nu}|_{C(t)}$ are $\rho$ \& $\vec{\nu}$ restricted to the
domain that satisfies $C(t)$.

In general,$(\rho_{C(t)},\vec{\nu}_{C(t)})$ will not satisfy Eqs.
\ref{eq: mp cont} \& \ref{eq: mp quant navier}. In particular, they
won't be satisfied if $C(t)$ reflects changes in information about
the system. None the less, if $(\rho,\vec{\nu})$ is a Schrodinger
distribution, then its time evolution together with $C(t)$ will yield
$(\rho_{C(t)},\vec{\nu}_{C(t)})$'s time evolution.

To be described by a Schrodinger distribution, or a conditional distribution
derived from a Schrodinger distribution, a system's environmental
state must contain a certain amount of information about the system.
For a quantum positional system, these distributions represent the
most information about the system's current state that its environment
can posses. The environment can, however, have less information. In
this case a system will be described by a density matrix distribution.
A density matrix distribution is of the form $\rho=\sum_{i}P_{i}\rho_{i}$,
$\vec{j}=\sum_{i}P_{i}\vec{j}_{i}$, where each $(\rho_{i},\vec{\nu}_{i})$
is a Schrodinger distribution or a conditional distribution, each
$P_{i}\geq0$, and $\sum_{i}P_{i}=1$. If the information about the
system contained in the environmental state remains constant, then
the $P_{i}$ are constant and the $(\rho_{i},\vec{\nu}_{i})$ Schrodinger/conditional
distributions evolve as usual.

\section{Types of quantum worlds}

Quantum theory has a great deal to say about what can be observed,
and about the probabilities of what might be observed, but it has
nothing to say about what exists. This is why there are a variety
of types of quantum worlds: types of worlds in which quantum theory
holds true, but in which different sorts of things exist, or the things
that exist follow different sorts of rules.

Quantum positional systems are one type of quantum world. Here, we
will review others types, together with their relationships to quantum
positional systems.

Because all types of quantum worlds obey quantum theory, they are
all observationally equivalent. There is therefore no way to know
which type corresponds to our world. Similarly, there is no point
in arguing which one ``correct''.

In spite of their observational equivalence, an understanding of types
of quantum worlds proves useful. One reason for this is that various
problems are easier to solve when described within a particular type
of quantum world. For example, any observable tied to tunneling time
may be easier to calculate using Bohmian mechanics, where tunneling
time can be directly calculated. In cases like this, types of quantum
worlds play a role akin to Lagrangian and Hamiltonian mechanics in
classical mechanics.

Types of quantum worlds also determine the full set of consequences
of the theory. For example, we know that the statement ``In quantum
mechanics, the electron in a double slit experiment travels through
both slits'' does not follow from quantum theory, because it does
not hold for quantum positional systems. Types also determine what
sorts of properties are consistent with quantum theory. As an example,
quantum positional systems tell us that we can self-consistently assume
that particle momentum means the same thing in quantum mechanics as
it does in classical mechanics. This does not follow from the theory,
because there are also quantum worlds in which it does not hold, but
it is consistent with the theory. In these cases, types of worlds
play a role in scientific theories that's akin to the one played by
models in mathematical theories.

For these reasons, it would be useful to have an understanding of
the range of types of quantum worlds.

\subsection{Bohmian mechanics}

In Bohmian mechanics (originally discovered by de Broglie, so perhaps
``de Broglian mechanics'' would be more appropriate) particles follow
paths with velocity $\vec{v}=\vec{j}/\left|\psi\right|^{2}$ where
$\vec{j}$ is the quantum current. This velocity is the same quantity
as $\vec{\nu}$, the velocity expectation value in quantum positional
systems. As is always the case for quantum worlds, the probability
for the initial particle position is given by $P(\vec{x})=\left|\psi(\vec{x})\right|^{2}$.
If the initial probability satisfies $P(\vec{x})=\left|\psi(\vec{x})\right|^{2}$,
and particles follow paths that satisfy $\vec{v}=\vec{j}/\left|\psi\right|^{2}$,
then $P(\vec{x},t)=\left|\psi(\vec{x},t)\right|^{2}$ holds for all
time. 

In Bohmian mechanics, position is the only true observable. Quantities
other than position are defined operationally, in terms of how they
are measured. The devices that measure these quantities can be designed
so that their outputs are distinguished by position (via a gauge reading,
ink on a printout, etc), in which case the measurement can be reduced
to a position measurement on the measuring device's Bohmian particles.

Bohmian mechanics is the limit of positional mechanics as $\varGamma\rightarrow\infty$.
In this limit, particles will follow paths with velocity $\vec{\nu}$
with probability $1$. The resulting dynamics are not positional,
because some of the properties of positional systems are violated;
in particular, a particle path's tangent vector is not generally equal
to the particle's linear momentum. The dynamics are, instead, Bohmian.

\subsection{Stochastic mechanics\label{subsec:Stochastic-mechanics}}

Stochastic mechanics is another type of quantum world that can be
generated from positional dynamics by taking a $\varGamma\rightarrow\infty$
limit. But in this case, the limit is taken differently than it is
for Bohmian mechanics. Start with a closed positional system with
some initial value of $\varGamma$, $\varGamma_{0}$. For every path
in the closed system's r-set, and for every particle in the system,
define $\triangle\vec{v}_{i}(t)=\vec{v}_{i}(t)-\vec{\nu}_{i}(t)$,
where $\vec{v}_{i}(t)$ is the particle's velocity at time $t$ \&
$\vec{\nu}_{i}(t)$ is the particle's mean velocity (given the state
of the closed system). When taking the limit as $\varGamma\rightarrow\infty$,
simultaneously scale each particle's instantaneous velocity as $\vec{v}_{i}(t)\rightarrow\vec{\nu}_{i}(t)+\sqrt{\frac{\varGamma}{\varGamma_{0}}}\triangle\vec{v}_{i}(t)$.
In this limit, the closed system executes a continuous random walk.
As the $\varGamma\rightarrow\infty$ limit decreases the jitter around
the mean due to the law of large numbers, the $\sqrt{\frac{\varGamma}{\varGamma_{0}}}$
in the expression $\vec{v}_{i}$ for increases it. These two offset,
leaving the system's randomness intact. 

Quantum stochastic models have previously been developed by Imre Fényes
\citep{Fenyes} and Edward Nelson \citep{Nelson 1}, among others.
They have shown that stochastic dynamics can reproduce significant
areas of quantum theory (see, for example, \citep{Nelson 2}). However,
the models described here are somewhat different from those earlier
models. In particular, Nelson was concerned with an error that occurred
within his stochastic theory. He considered two entangled but decoupled
harmonic oscillators\citep{Nelson 2}. Quantum mechanically, the position
correlation between two such oscillators does not decay as $t\rightarrow\infty$.
In Nelson's stochastic model, on the other hand, the correlation does
decay to $0$ as $t\rightarrow\infty$. In contrast to Nelson, and
in agreement with quantum theory, the stochastic limit used here does
not decay. This is because, unlike Nelson's model, here it is the
path of the closed system as a whole that executes the random walk.
When concatenation chooses a new path, the individual particle paths
are not chosen independently.

It is likely clear that a world that satisfies stochastic mechanics
as defined here will satisfy quantum theory in much the same manner
that Bohmian mechanics does.

\subsection{Many worlds interpretation}

In the many worlds interpretation (``mwi''), a type of quantum world
is crafted using only those things that are directly referenced within
quantum theory. It can be derived from quantum positional systems,
in that quantum theory can be derived from quantum positional systems.
However, unlike the prior two types of quantum worlds, what exists
in an mwi world is not derived from what exists in quantum positional
systems. If a world has to be build out of only those things that
are referred to in quantum theory, the only option is to grant existence
to the wave function.

Therefore in an mwi world only a great, universal wave exists. This
wave resides in a space of particle configurations; for concreteness,
we'll stick to the position basis. Each configuration consists of
a set of particles, with each particle at a particular position. Because
the wave function does not exclude any part of the physical reality,
each configuration can be viewed as being a complete universe. The
space that the wave inhabits is the space, or ``multiverse'', that
contains all of these universes. The universe we see all around us
would be one of the universes in a multiverse.

Given that $\left|\Psi\right\rangle $ is the multiverse's wavefunction
at some time, the probability that some statement $\alpha$ about
observable quantities holds true for our universe at that time is
$\left\langle \Psi\right|\mathbb{P}_{\alpha}\left|\Psi\right\rangle $.
(The same is true for any universe.) With that, let's say that instead
of being given the universal wave-function $\Psi$, we are instead
given that $\beta$ is true in our universe. What is the probability
that $\alpha$ is also true? We can deduce $P(\alpha|\beta)$ if we
assume (1) the probability, when restricted to the wavefunctions of
any particular basis, is the same for all bases, and (2) a priori
equal probability. With these two assumptions, and the probability
law given above, it follows that $P(\alpha|\beta)=\frac{1}{Tr(\mathbb{P}_{\beta})}Tr(\mathbb{P}_{\beta}\mathbb{P}_{\alpha})$.

Now let's say the current time is $t_{1}$, and $\beta$ is ``the
current state implies that $\gamma$ was true at time $t_{0}$''.
In this case, at $t_{1}$, $\mathbb{P}_{\beta}=U(t_{1}-t_{0})\mathbb{P}_{\gamma}U^{\dagger}(t_{1}-t_{0})$.
Plugging this into the probability rule, $P(\alpha,t_{1}|\beta,t_{1})=P(\alpha,t_{1}|\gamma,t_{0})=\frac{1}{Tr(\mathbb{P}_{\gamma})}Tr(U(t_{1}-t_{0})\mathbb{P}_{\gamma}U^{\dagger}(t_{1}-t_{0})\mathbb{P}_{\alpha})$,
as usual.

Implicitly, this means that within an mwi multiverse, the interpretation
of these probabilities is more restrictive than we are used to. Informally,
the statement ``Given that it is cloudy in the morning, the probability
that it will rain in the afternoon is 50\%'' can be given two equivalent
interpretations. One (less commonly stated) is: Given that it is the
afternoon, and we're in a physical state that contains the information
that it was cloudy in the morning, the probability that it is raining
right now is 50\%. This interpretation is well defined within an mwi
multiverse. The second (more common) one is: Given that it is the
morning, and it is cloudy outside, the probability that it will rain
in the afternoon is 50\%. This one is \emph{not} well defined within
an mwi multiverse, because the initial state will generally evolve
into multiple futures, in some of which it is raining, and in some
of which it is not.

However, in any real world application of probabilities, the first
(more cumbersome) interpretation may always be used. It follows that
the many worlds interpretation reproduces all the observable qualities
of quantum theory, and so is a type of quantum world.

(Side note: Within quantum theory, the notion of an observable state
containing information about the past can be somewhat subtle. It is
analyzed at length in \citep{Tesse}.)

\subsection{Physical collapse}

It seems likely that a type of quantum world can be crafted that is
like the many worlds interpretation, but that eliminates universes
we know not to be ours. This type is described by physical collapse
models.

Consider an experiment that gathers information about some system.
In any experimental run, the observed state of the measuring devices
will change with time, ultimately settling on their final states.
Using quantum theory, one could derive the possible time sequences
of measuring device observations ending in each possible final state,
along with their probabilities. That is, one could obtain a stochastic
equation of motion for these quantities. One could then construct
the conditional wave function for the closed system, conditioned on
the values of the measuring devices as a function of time. This will
yield a stochastic equation of motion for the wave function. Such
an equation of motion would combine both the Schrodinger and information
gathered about our universe in a single equation.

This is what physical collapse theories achieve. A number of such
theories currently exist (see, for example, \citep{GRW} \& \citep{CSL}).
In Tilloy \& Wiseman (2021) \citep{Wiseman} it was shown that these
various physical collapse theories can be produced by coupling an
otherwise closed quantum system to a bath of Bohmian particles. Conditioning
the quantum system's wave function on the position of the Bohmian
particles yields the required stochastic equations of motion. As the
Bohmian particles can be replaced by positional particles, physical
collapse models can be wholly derived from positional systems.

\section{\label{sec:Relativistic-Amendments}Relativistic Amendments}

Here, a relativistic version of positional systems will be briefly
presented. The primary purpose will be to demonstrate that positional
dynamics can be made relativistically covariant. There will be no
attempt to hone the assumptions to fully replicate relativistic quantum
field theory.

\subsection{Particle states \& paths}

Relativistically, a system's state can be described along any space-like
surface. For positional systems, the system state on any given space-like
surface continues to be comprised of its particles' positions \& identifying
characteristics along the surface, together with the field state on
the surface. Field states will be discussed in the final section.

Particle paths remain continuous and piecewise differentiable, and
a particle's 4-mometum is its mass times its path's tangent 4-vector.
In a relativistic r-set, a system's 4-momentum is conserved over any
period during which it does not interact with its environment. The
full set of paths remains the closure of the r-set under concatenation.
However, relativistically, concatenations \& interactions can no longer
occur along time slices. Instead, the means for constructing the space-like
hyper-surfaces along which these events take place has to be frame
independent. It should also reproduce the non-relativistic description
in the required limit.

There are a number of ways of accomplishing this. One of them will
be described below. Briefly: Choose a surface whose normal at each
point is the expected mass 4-current. To define what ``the expected
mass 4-current at each point'' means, we will first briefly pause
to describe the relativistic joint probability and 4-current probability.

\subsection{Relativistic transformation of $\eta$}

As the system state can be defined along any space-like hyper-surface,
and $\eta$ can be conditioned on any portion of the system state,
$\eta$ can be conditioned on the state along any space-like hyper-surface.
$\eta$ can also be expressed in any coordinate system. The coordinate
system does not need to be related to the space-like hyper-surface(s)
on which $\eta$ is conditioned.

Here we will consider the $\eta$ for a single particle. The information
contained in condition is assumed to be fixed, and $\eta(\tilde{x},\tilde{u})$
is known for all 4-space positions, $\tilde{x}$, and all 4-velocites,
$\tilde{u}$, given the condition. It will be shown shortly that $\tilde{\eta}=\left(\eta(\tilde{x},\tilde{u}),\vec{v}\eta(\tilde{x},\tilde{u})\right)$
is a 4-vector, where $\vec{v}=\nicefrac{\vec{u}}{u_{0}}.$ This allows
us to write $\eta(\tilde{x},\tilde{u})$ along any space-like hyper-surface:
If $\tilde{s}(\tilde{x})$ is the normal to $S$ at point $\tilde{x}$,
$\eta_{S}(\tilde{x},\tilde{u})=\tilde{s}(\tilde{x})\cdot\tilde{\eta}(\tilde{x},\tilde{u})=\frac{\tilde{s}(\tilde{x})\cdot\tilde{u}}{u_{0}}\eta(\tilde{x},\tilde{u})$.

\emph{Sketch of proof that $\left(\eta(\tilde{x},\tilde{u}),\vec{v}\eta(\tilde{x},\tilde{u})\right)$
is a 4-vector}: To Lorentz transform $\eta$, consider any small spacial
volume in the current frame, $\triangle V$, and any flat space-like
hyper-surfaces that intersects $\triangle V$. Take $\triangle V$
to be small enough so that particle paths between $\triangle V$ and
the intersecting hyper-surface can be well approximated by free paths.
For any 4-velocity, $\tilde{u}$, define $\triangle V_{\tilde{u}}^{\prime}$
to be the volume of the space-like hyper-surface crossed by paths
that cross $\triangle V$ with 4-velocity $\tilde{u}$ . 

Take $\tilde{x}$ to be any point on the intersection of $\triangle V$
and the space-like hyper-surface. $\eta$ transforms by the rule:
$\eta(\tilde{x},\tilde{u})\triangle V=\eta^{\prime}(\tilde{x},\tilde{u})\triangle V_{\tilde{u}}^{\prime}$.
A simple calculation yields: $\eta^{\prime}(\tilde{x},\tilde{u})=\frac{1}{\sqrt{1-v_{r}^{2}}}\left(1-\vec{v}_{r}\cdot\vec{v}\right)\eta(\tilde{x},\tilde{u})$,
where $\vec{v}_{r}$ is the relative velocity of the two frames. This
is precisely the manner in which $u_{0}$ transforms, $u_{0}^{\prime}=\frac{1}{\sqrt{1-v_{r}^{2}}}\left(1-\vec{v}_{r}\cdot\vec{v}\right)u_{0}$,
meaning that $\frac{1}{u_{0}}\eta(\tilde{x},\tilde{u})$ is a scalar.
$\left(u_{0}\frac{1}{u_{0}}\eta(\tilde{x},\tilde{u}),\vec{u}\frac{1}{u_{0}}\eta(\tilde{x},\tilde{u})\right)=\left(\eta(\tilde{x},\tilde{u}),\vec{v}\eta(\tilde{x},\tilde{u})\right)$
is therefore a 4-vector, $\tilde{\eta}$.

\subsection{The probability current \& the 4-velocity distribution}

The relativistic probability 4-current, $\tilde{j}$, is defined as
$\tilde{j}=\left(\rho,\vec{j}\right)=\intop\tilde{\eta}(\tilde{x},\tilde{u})d\tilde{u}$.
Unlike the non-relativistic case, the relativistic probability current
\& momentum are not closely interrelated, because the probability
current is related to velocity, while momentum is related to the 4-velocity.

The 4-velocity distribution at $\tilde{x}$ is $\upsilon(\tilde{u}|\tilde{x})=\frac{1}{\rho(\tilde{x})}\tilde{\eta}(\tilde{x},\tilde{u})$.
Along some other flat space-like hyper-surface, $S$ with normal $\tilde{s}$,
the position distribution is $\rho_{S}(\tilde{x})=\tilde{s}\cdot\tilde{j}(\tilde{x})$,
and $\eta_{S}(\tilde{x},\tilde{u})=\tilde{s}\cdot\tilde{\eta}(\tilde{x},\tilde{u})=\rho_{S}(\tilde{x})\upsilon_{S}(\tilde{u}|\tilde{x})$.
It follows that $\upsilon_{S}(\tilde{u}|\tilde{x})=\frac{1}{\tilde{s}\cdot\tilde{j}}\tilde{s}\cdot\tilde{\eta}=\frac{1-\vec{v}\cdot\vec{N}}{1-\vec{\nu}\cdot\vec{N}}\upsilon(\tilde{u}|\tilde{x})$,
where $\vec{N}=\frac{\vec{s}}{s_{0}}$.

The local 4-velocity probability distribution, $\upsilon(\tilde{u}|\tilde{x})$,
is therefore not a scalar. This is because of the manner in which
$\eta(\tilde{x},\tilde{u})$ transforms. $\triangle V_{\tilde{u}}^{\prime}$
may shrink or attenuate by different amounts, depending on $\tilde{u}$.
As a result, different $\tilde{u}$ will make differing contributions
to $\upsilon_{S}(\tilde{u}|\tilde{x})$ than they did to $\upsilon(\tilde{u}|\tilde{x})$.

\subsection{Field conditioned $\eta$}

There will be a particular interest in $\eta$ conditioned on the
field state alone. Classically, the electromagnetic field state contains
an enormous amount of information about particle states \& paths.
When the field is discretized, we expect information about particle
state \& path to be degraded; it yields a probability distribution,
$\eta$, rather than the actual values.

In general, the discrete field state also can not be expected to uniquely
specify the number or types of particles. If we represent a fixed
collection of particles by a tuple of the particles' identifying characteristics,
the field state will specify that the number \& types of particles
must some element in a collection of such tuples, $\Lambda$. The
field conditioned particle distribution can then be written:

\[
\eta=\sum_{\vec{\iota}\in\Lambda}P(\vec{\iota})\eta_{\vec{\iota}}(\tilde{x}_{1},\tilde{u}_{1},\tilde{x}_{2},\tilde{u}_{2},...)
\]

Where $P(\vec{\iota})$ is the probability that the number \& types
of particles is described by $\vec{\iota}$.

{[}Side Note: A curious aspect of relativistic quantum theory is that
position is not an observable quantity. Conditioning on the field
state allows for a simple physical explanation of this. Classically,
particle states can be mixed and matched in any fashion, so the state
of any set of particles doesn't reveal anything about the other particles;
information about outside particles is only available through the
field state. Let's assume this is also the case for quantum systems.
With that, if the information stored within quantum fields does not
allow us to conclude with absolute certainty that any particle must
be within any limited region of space, then position is lost as a
quantum observable. Because position can be determined to extreme
high probability, position can still be something that we observe
(e.g., our brain's field state can be interpreted as implying the
position of various objects), but position can not be represented
within relativistic quantum theory as an observable quantity.

Within quantum theory, this still leaves the technical issue of the
nature of $\psi^{\dagger}(\tilde{x})\psi(\tilde{x})$'s domain. Relativistically,
it can't be where the particle will be observed, as position is not
an observable. Within positional dynamics, the answer is the obvious
one: This the probability distribution of where the particle is, given
the field state.{]}

We can now return to the question of the surfaces along which momentum
changing events take place.

\subsection{Cut surfaces}

Let's refer to a surface along which a momentum changing event takes
place as a ``cut''. Let's assume that such an event has taken place.
We need to be able to construct the set of possible surfaces along
which the next one can take place. There are any number of ways to
do this. Here, one such method will be described.

Start with $\eta$ on the prior cut, conditioned on the field state
on the cut: $\eta=\sum_{\vec{\iota}\in\Lambda}P(\vec{\iota})\eta_{\vec{\iota}}(\tilde{x}_{1},\tilde{u}_{1},\tilde{x}_{2},\tilde{u}_{2},...)$.

For each $\vec{\iota}\in\Lambda$, each particle in $\vec{\iota}$,
extract the single particle joint probability distribution: $\eta_{\vec{\iota},i}(\tilde{x},\tilde{u})=\int\eta_{\vec{\iota}}(\tilde{x}_{1},\tilde{u}_{1},\tilde{x}_{2},\tilde{u}_{2},...)\Pi_{j\neq i}d\tilde{u}_{j}\Pi_{j\neq i}d\tilde{x}_{j}$

Propagate each $\eta_{\vec{\iota},i}$ through the region following
the cut using the free propagator, $\boxempty\cdot\tilde{\eta}_{\vec{\iota},i}=0$.

Calculate the individual particle mass current in the region following
the cut: $m_{\vec{\iota},i}(\tilde{x})=m_{i}\int\tilde{n}_{\vec{\iota},i}(\tilde{x},\tilde{u})d\tilde{u}$

Calculate the total mass current: $\tilde{m}(\tilde{x})=\sum_{\vec{\iota}\in\Lambda}P(\vec{\iota})\sum_{i\in\vec{\iota}}m_{\vec{\iota},i}(\tilde{x})$.

Normalize the vectors in the field to $1$: $\tilde{s}(\tilde{x})=\frac{1}{\left|\tilde{m(}\tilde{x})\right|}\tilde{m}(\tilde{x})$.

A surface is a possible cut surface if $\tilde{s}(\tilde{x})$ is
the surface's normal at all points on the surface. 

For the world as we experience it, such a surface will never wander
far from what we view as a time slice. If there is an inhabited world
moving at near light speed away from us, in the vicinity of that world
the cut surface will also never stray far from what those inhabitants
would view as a time slice. This is required for the relativistic
case to reduce to the non-relativistic one. 

For an isolated particle, $\tilde{s}(\tilde{x})$ is the 4-velocity
associated with $\vec{\nu}=\frac{1}{\rho}\vec{j}$. The means of constructing
the cut described here has been chosen so that this would be the case.
More generally, for any multiparticle wavefunction, $\tilde{s}(\tilde{x})$
can be calculated from the wavefunction's probability current and
the particle masses. 

There is, however, a slight complication. Along the prior cut, $\tilde{s}(\tilde{x})$
will likely be quite close to the normal to that cut, $\tilde{s}_{c}(\tilde{x})$,
but they may not be equal. This is because the field state can discontinuously
change on the cut, causing the two to be different. As a result, a
surface constructed as described can cross the prior cut. This should
not be allowed. There are two ways to eliminate the possibility: either
eliminate any surface that crosses the prior cut, or modify $\tilde{s}(\tilde{x})$
so that as the vector field approaches the prior cut, it smoothly
transitions from $\tilde{s}(\tilde{x})$ to $\tilde{s}_{c}(\tilde{x})$.
In the second case, $\tilde{s}(\tilde{x})$ as initially described
will remain a good approximation.

\subsection{Equation of motion for $\eta$}

The relativistic equation of motion for the single particle \noun{$\eta$},
under conditions of constant information, can be derived in much the
same manner as the non-relativistic equation of motion. Here, we'll
take the field of cut normals, $\tilde{s}(\vec{x},t)$, to be given.
If the field of cut normals is not knowable given the state of the
particle's environment, then we will have to average the $\tilde{\eta}$
derived here over the space of possible cut fields.

Choose any point $\left(\vec{x},t\right)$, and any region of $\left(\vec{x},t\right)$'s
backward light-cone small enough so that within it the cut normals,
$\tilde{s}$, are approximately constant. Choose the terminal surface
to the light cone volume to have $\tilde{s}$ as its normal, so any
cut will slice through the whole volume. The distance in the $\tilde{s}$
direction from 4-volume's terminal $\tilde{s}$-surface to $\left(\vec{x},t\right)$,
$\triangle\tau$, should be small. 

With this set-up, a calculation nearly identical to the non-relativistic
one yields the equation of motion:

\[
\boxempty\cdot\tilde{\eta}=\varGamma\tilde{s}(\vec{x},t)\cdot\left(\tilde{j}(\vec{x},t)q(\tilde{u}|\vec{x},\iota,e_{S})-\tilde{\eta}(\vec{x},t,\tilde{u})\right)
\]

Where $\varGamma$ and $q(\tilde{u}|\vec{x},\iota,e_{S})$ are as
before: $\varGamma$ is the expected number of momentum changing events
per unit time (as measured in the $\tilde{s}$ direction), and $q(\tilde{u}|\vec{x},\iota,e_{S})$
is the 4-velocity probability distribution upon a momentum changing
event, given the state along the cut. The right hand side can be written:
$\left(\tilde{s}(x,t)\cdot\tilde{j}(\vec{x},t)\right)\times\varGamma\left(q(\tilde{u}|\vec{x},\iota,e_{S})-\upsilon_{S}(\tilde{u}|\tilde{x})\right)$;
as $\varGamma$ gets large, $q(\tilde{u}|\vec{x},\iota,e_{S})-\upsilon_{S}(\tilde{u}|\tilde{x})$
grows small.

Integration over $\tilde{u}$ yields $\boxempty\cdot\tilde{j}=0$.

If the means for constructing cut surfaces is as described in the
prior section, then in regions where the particle is isolated $\tilde{s}(x,t)=\frac{1}{\left|\tilde{j}\right|}\tilde{j}$.
With $\mathbb{j}=\frac{1}{\left|\tilde{j}\right|}\left(\rho-\vec{v}\cdot\vec{j}\right)$,
the equation of motion is then $\boxempty\cdot\tilde{\eta}=\varGamma\left(\left|\tilde{j}\right|q(\tilde{u}|\vec{x},\iota,e_{S})-\mathbb{j}\eta(\vec{x},t,\tilde{u})\right)$.

\subsection{Field states}

Relativistically, particle interactions take place across light-like
separations. This, together with momentum conservation in the r-set,
means that some element of the system path must carry the interaction
4-momentum between the time it is given, and the time it is received.
That element is the field state. 

In positional systems, these interactions take place discretely. The
discrete elements of the field state will be called ``signals''.
Each signal carries 4-momentum, the momentum being an element of the
signal's state. During a particle-field interaction, the particle
will absorb a set of signals, which are thereby removed from the field
state, and emit a set of signals, which are added to the field state.
One of these sets of signals may be empty. In the r-set, the sum of
the particle's initial momentum \& the total momentum of the absorbed
signals must equal the sum of the particle's final momentum \& the
total momentum of the emitted signals. The particle's mass may be
assumed to remain unchanged following the interaction.

This description differs somewhat from QED, because QED only has a
3-point interactive vertex. The vertices mentioned here may be viewed
as being higher order vertices generated by QED, but here taken to
be execute in an instant. These higher order vertices will generally
encompass multiple particles interacting with the field along the
cut.

Other elements may be added to the signal state, beyond the 4-momentum.
For example, polarization. If more than one type of interactive field
is being considered, identifying characteristics can also be added. 

The field state is largely responsible for generating the artifacts
utilized in positional mechanics, such as $q$, which governs interactions,
$\eta$, which encapsulates our information about particles, and spin,
which contains aspects of both. Practically speaking, we also deduce
the field state based largely on the values of those artifacts.

All of these artifacts are functions of particle position. Deriving
them can be enabled by transforming the field state to a position
representation. For quantum theory, this is accomplished via Fourier
transformations.

However it is accomplished, the position representation of the field
state allows the field state to be described along any space-like
hyper-surface. This description ought to be accurate, meaning that
when the position representation along the hyper-surface is transformed
back to the signal state representation, it yields the set of signals
that would actually exist on the hyper-surface. Which is to say, the
set of signals is consistent with the 4-space locations at which particles
absorbed and/or emitted signals.


\begin{thebibliography}{10}
\bibitem{Wigner}Wigner, E. P. (1932). \emph{On the quantum correction
for thermodynamic equilibrium}, Physical Review. 40 (5): 749--759.

\bibitem{Everett}Hugh Everett (1957). \emph{Relative State Formulation
of Quantum Mechanics}, Reviews of Modern Physics 29: 454-462.

\bibitem{Bohm 1}Bohm, David (1952). \emph{A Suggested Interpretation
of the Quantum Theory in Terms of \textquotedbl Hidden Variables\textquotedbl{}
I}, Physical Review. 85 (2): 166--179.

\bibitem{Bohm 2}Bohm, David (1952).\emph{ A Suggested Interpretation
of the Quantum Theory in Terms of \textquotedbl Hidden Variables\textquotedbl{}
II}, Physical Review. 85 (2): 180--193.

\bibitem{Durr overview}Durr, D.; Zanghi, N.; Goldstein, S. (Nov 14,
1995). \emph{Bohmian Mechanics as the Foundation of Quantum Mechanics},
\href{https://arxiv.org/abs/quant-ph/9511016}{arXiv:quant-ph/9511016}

\bibitem{Holland overview} Peter R. Holland (1993). \emph{The Quantum
Theory of Motion: An Account of the De Broglie-Bohm Causal Interpretation
of Quantum Mechanics}, Cambridge University Press, ISBN 0-521-35404-8

\bibitem{Fenyes}Fényes, I. (1946). \emph{A Deduction of Schrödinger
Equation}, Acta Bolyaiana. 1 (5): ch. 2.

\bibitem{Nelson 1}Nelson, Edward (1966). \emph{Derivation of the
Schrödinger Equation from Newtonian Mechanics}, Physical Review. 150
(4): 1079--1085.

\bibitem{Nelson 2}Edward Nelson (2012). \emph{Review of Stochastic
Dynamics}, J. Phys.: Conf. Ser. 361 01201.

\bibitem{Madelung}E. Madelung (1926), \emph{Eine anschauliche Deutung
der Gleichung von Schrödinger}, Naturwissenschaften 14: 1004-1004

\bibitem{Bell 1}Bell, John S. (1982). \emph{On the Impossible Pilot
Wave}, Foundations of Physics, 12(10): 989--999. Reprinted in Bell
1987c: 159--168.

\bibitem{Shucker}Shucker, D. (1980). \emph{Stochastic Mechanics of
Systems with Zero Potential}, J. Functional Analysis 38: 146-155.

\bibitem{Ghirard}G. C. Ghirard, C. Omero, A. Rimini, I. Weber (1978).
\emph{The stochastic interpretation of quantum mechanics: A critical
review}, Riv. Nuovo Cim. 1: 1--34.

\bibitem{Bell 2}Bell, J.S. (1964). \emph{On the Einstein-Podolsky-Rosen
paradox}, Physics, 1: 195--200; reprinted in Bell 1987b {[}2004{]}:
14--21.

\bibitem{Kochen}S. Kochen; E. P. Specker (1967). \emph{The problem
of hidden variables in quantum mechanics}, Journal of Mathematics
and Mechanics, 17 (1): 59--87.

\bibitem{Norsen}Travis Norsen (2014). \emph{The Pilot-Wave Perspective
on Spin}, American Journal of Physics 82: 337-348.

\bibitem{PBR}Pusey, Matthew F.; Barrett, Jonathan; Rudolph, Terry
(2011). \emph{On the reality of the quantum state}, \href{https://arxiv.org/abs/1111.3328}{arXiv:1111.3328 [quant-ph]}

\bibitem{Tesse}Tesse, E. (2026). \emph{The trouble with recoding
devices} \href{https://arxiv.org/abs/2602.02191}{arXiv:2602.02191 [quant-ph]}

\bibitem{Wiseman}Antoine Tilley; Howard Wiseman (2021). \emph{Non-Markovian
wave-function collapse models are Bohmian-like theories in disguise},
\href{https://arxiv.org/abs/2105.06115}{arXiv:2105.06115 [quant-ph]}

\bibitem{GRW}Ghirardi, G.C.; Rimini, A.; Weber, T. (1986). \emph{Unified
dynamics for microscopic and macroscopic systems}, Physical Review
D. 34 (2): 470--491.

\bibitem{CSL}Pearle, Philip (1989). \emph{Combining stochastic dynamical
state-vector reduction with spontaneous localization}, Physical Review
A. 39 (5): 2277--2289.

\bibitem{Durr 2}Detlef Dürr , Sheldon Goldstein, and Nino Zanghì
(1992), \emph{Quantum Equilibrium and the Origin of Absolute Uncertainty},
Journal of Statistical Physics, 67(5): 843--907.

\bibitem{Fabbri} Fabbri, Luca (2025). \emph{Madelung structure of
the Dirac equation}, J. Phys. A: Math. Theor. 58: 195301. \href{https://arxiv.org/abs/2505.05861}{ arXiv:2505.05861}
\end{thebibliography}
\end{document}